\documentclass[12pt]{article}
\usepackage{graphicx}
\usepackage{subfigure}
\usepackage{amsmath}
\usepackage[psamsfonts]{amssymb}
\usepackage{mathrsfs}
\usepackage{amsthm}
\usepackage{xspace}
\usepackage{array,color}
\usepackage{graphicx}
\def\baselinestretch{1.1}
\makeatletter\@addtoreset{equation}{section}\makeatother

\newcommand{\be}{\begin{equation}}
\newcommand{\ee}{\end{equation}}
\newcommand{\bea}{\begin{eqnarray}}
\newcommand{\eea}{\end{eqnarray}}

\def\bsp{\be\begin{split}}

\def\inc#1#2{{%
\raise-.5\ht1\box1%
}}

\renewcommand{\title}[1]{\vbox{\center\LARGE{#1}}\vspace{5mm}}
\renewcommand{\author}[1]{\vbox{\center#1}\vspace{5mm}}
\newcommand{\address}[1]{\vbox{\center\em#1}}
\newcommand{\email}[1]{\vbox{\center\tt#1}\vspace{5mm}}

\begin{document}
\begin{titlepage}
\begin{center}
\vspace{5mm}
\hfill {\tt }\\
\vspace{20mm}
\title{From Liouville to Chern-Simons, Alternative Realization of Wilson Loop Operators in AGT
Duality} \vspace{10mm}
\author{\large Jian-Feng Wu\footnote{wujf@itp.ac.cn}, Yang Zhou\footnote{yzhou@itp.ac.cn}}

\address{Institute of Theoretical Physics\\
Chinese Academy of Sciences, Beijing 100190, PRC}

\email{wujf@itp.ac.cn, yzhou@itp.ac.cn}

\end{center}
\vspace{10mm}

\abstract{ \noindent We propose an $SL(2,\mathbb{R})$ Chern-Simons
description of Liouville field theory (LFT), whose correlation
function duals to partition function of $\mathcal{N}=2\,\,\, SU(2)$
gauge theories. We give the dual expressions for conformal blocks,
fusion rules, and Wilson loop operators. By realizing Wilson loop
operator in Liouville as a Hopf link in $S^3$ on which lives an
$SL(2,\mathbb{R})$ Chern-Simons theory, we obtain an alternative
description of monodromy of this loop operator in Liouville field
theory as the ratio of link invariants. We show how to calculate
t'Hooft loops in the simplest example -- the $\mathcal{N}=4$ super
Yang-Mills theory. The results we obtained are consistent with those
in 0909.0945 and 0909.1105.}

\vfill

\end{titlepage}

{\addtolength{\parskip}{-1ex}\tableofcontents}

\section{Introduction}

Recently, Alday, Gaiotto and Tachikawa (AGT)~\cite{AGT} established
a new duality between Liouville theory and four dimensional
$\mathcal{N}=2$ gauge theories. These $\mathcal{N}=2$ gauge theories
can be obtained by compactifying coincided $M5$ branes on specified
Reimann surfaces $C$ with punctures~\cite{Gaiotto}, which builds a
bridge between 2d and 4d field theories, thus theories living on one
side will obtain new information and then benefit from the other.
Later after that, many people generalized this duality to various
situations~\cite{Wyllard,Mironov etal}. Also there are some
developments on this new duality from different
features~\cite{Nanopoulos,Benini1,Benini2,Maruyoshi,Gaiotto
0908,Iguri Nunez,Nekrasov
Shatashvili,Jorjadze,Poghossian,Bonelli,Giombi
Pestun,Alday,Gadde,Yang Zhou,Yamada,Papadodimas}. Dijkgraaf and
Vafa~\cite{DV} even proved this duality at a very general level in
two different approaches. In~\cite{AGT}, it was shown that the
Nekrasov partition
 functions~\cite{Nekrasov1,Nekrasov2,Nekrasov:2004vw} of the generalized quiver gauge
theories on $\mathbb{R}^4$ are identical to the correlation functions in Liouville field theory. In this duality, the
Liouville momenta at the marked points specify the masses of the flavor multiplets, while the momenta in the
intermediate channels are identified as the Coulomb branch parameters. Very recently, two groups (Gaiotto et
al~\cite{AGGTV} and Drukker et al~\cite{DGOT}) calculated Wilson loop operators in gauge theories using the dual
Liouville language, the results perfectly match with those in gauge theories. This can also be seen as a strong test of
the AGT duality.

Loop operators in $\mathcal{N}=2 \,\,\,SU(2)$ 4d quiver gauge
theories are believed to be monodromies of related conformal blocks
in the dual Liouville theory. To be precise, a loop operator in
$\mathcal{N}=2\,\,\, SU(2)$ gauge theories can be identified, in the
Liouville theory side, as an insertion of Liouville loop operator in
related conformal blocks of correlation functions which involve two
chiral degenerated primary operators. In order to obtain the loop
operator, we need a long complicated calculation in which fusion and
braiding matrices are highly involved. To simplify this, we noticed
that one can always realize a loop operator on Riemann suface $C$ as
a world line of a charged particle and then gluing the time
direction back to its origin. This actually visualizes the loop
operator as a knot or link in the extended world-volume of the
Riemann surface $C$. Using the standard surgery method, one can
always glue this world-volume as $S^3$, the eigenvalue of a loop
operator acting on a specific conformal block is identified to a
correlation function of a related knot or link in $S^3$ on which
there is a topological invariant theory. This surgery method is very
similar with that used in Chern-Simons-Witten(CSW)~\cite{CSW}
theory. The CSW theory is a dual geometric description of
Wess-Zumino-Witten(WZW) models which are rational conformal field
theories(RCFT)~\cite{Moore Seiberg}. We propose that the 3D dual
theory is exactly a Chern-Simons theory with gauge group
$SL(2,\mathbb{R})$. This conjecture, however, is not arbitrary since
there is a alternative realization of Liouville theory as a gauged
$SL(2,\mathbb{R})$ WZW model~\cite{Hamiltonian Reduction,Ishibashi,
Izawa}, which is believed to dual to an $SL(2,\mathbb{R})$
Chern-Simons theory~\cite{CSW;nocompact,Carlip 91,Banados 94,Carlip
05,Fjelstad Hwang}. In another way, to understand the physical
origin of the Liouville filed on the modular geometry, we expect
that this Liouville theory comes from a mother theory which has M
branes source, like a Chern-Simons type theory of M2
branes~\cite{BLG, ABJM}. Even though we have not known all details
of the modular theory, we will give a direct derivation from
Chern-Simons to Liouville theory as a realization on this proposal.
The derivation is slightly different from the well known CSW-WZW
correspondence. Surprisingly, the ingredients of Liouville theory
have very simple expressions in Chern-Simons theory side. The
holomorphic part of Liouville correlation function has been related
to knot invariant; fusion rules correspond to loop algebras;
braiding becomes passing through of two lines; sewing of conformal
blocks becomes surgery operation; different normalization of
Liouville partition function corresponds to different surgeries.
These correspondences offer enough tools to deal with loop
operators.

Using this proposal, realizing Wilson loop operator in Liouville as
a Hopf link in $S^3$, we obtain dual descriptions of these
monodromies in $SL(2,\mathbb{R})$ Chern-Simons theory which
completely depend on the modular properties of the related affine
algebra. They are ratios of corresponding correlation functions of
links (links invariants) in three dimension, which can be easily
calculated by standard surgery method. Now we have not to know
details about monodromy matrices and braiding factors. The only
ingredients are the modular S matrices of the affine algebra. This
method simplifies the calculation a lot. In general, all monodromies
which related to general loop operators have dual descriptions as
ratios of links invariants in $S^3$ with an $SL(2,\mathbb{R})$
Chern-Simons action. If this conjecture is valid in more general
situations, then one could expect this could be easily used in
$\mathcal{N}=2$ $SU(N)$ gauge theories and Toda field theories.
Thus, the problem refers to the modular S matrices and their
analytic continuations of characters of affine $sl(n)$ algebra.
There remain further works on it.

The structure of the paper is as following. In section 2, we reviewed some important results for AGT duality and loop
operators in this duality, including the dictionary, loops in gauge theory, loops in liouville theory. We also reviewed
the modular bootstrap for Liouville field theory in this section. In section 3, we showed that from an
$SL(2,\mathbb{R})$ Chern-Simons on a manifold without boundary, one can extract a gauged $SL(2,\mathbb{R})$ WZW model
which is exactly equivalent to a Liouville theory at the level of partition function. In section 4, we gave the
equivalent relations for Liouville ingredients and we calculated Wilson loops in general case and checked the simplest
t'Hooft loop. Section 5 leaves the conclusions.
\section{Loop Operators in AGT Duality}
In this section we review the relation between loop operators in $\mathcal{N}=2$ gauge theories and Liouville theory
briefly. For details, see original works by Gaiotto et al~\cite{AGT}~\cite{AGGTV} and Drukker et al~\cite{DGOT}.
\subsection{AGT Duality}
It was shown in~\cite{Gaiotto} a large class of four dimensional $\mathcal{N}=2$ $SU(2)$ SCFTs can be obtained by
compactifying the six-dimensional $(2,0)$ theory of type $A_1$ on a Riemann surface with punctures. Each puncture is
associated to an $SU(2)$ flavor symmetry, which can be used to give mass to the hypermultiplets. Each SCFT in this
class can be labeled by two integers $g,n,$ which are the genus and the number of punctures of the Riemann surface
$C_{g,n}$. The parameter space of the theory coincides with the complex moduli space of the punctured Riemann surface.

\begin{table}[!hbp]
\begin{center}
\begin{tabular}{|c|c|}
\hline \hline
\textbf{Gauge Theory} & \textbf{Liouville Theory}\\
\hline
                                                    &     Liouville parameters\\
Deformation parameters $\epsilon_1$, $\epsilon_2$   &   $\epsilon_1 : \epsilon_2 = b : 1/b$\\
                                                    &   $c=1+6Q^2, Q = b +1/b$\\
\hline
Mass parameter $m$              & Insertion of a  \\
associated to an $SU(2)$ flavor & Liouville vortex operator
$e^{2m\phi}$\\
 \hline
one $SU(2)$ gauge group & a thin channel with  \\
with UV coupling $\tau$ & sewing parameter $q=exp(2\pi i \tau)$\\
\hline
 Vacuum expectation value $a$ & Primary $e^{2\alpha \phi}$ for
the channel\\
of an $SU(2)$ gauge group     & $\alpha = Q/2 + a$ \\
\hline
Instanton part of $Z$ & Conformal blocks  \\
\hline
One-loop part of $Z$ & Product of DOZZ factors  \\
\hline
Integral of   $|Z^{2}_{full}|^2$ & Liouville correlator \\
\hline
\end{tabular}
\end{center}
\caption{Dictionary between the Liouville correlation functions and
Nekrasov's partition function $Z$. This table comes directly from
\cite{AGT}.}
\end{table}

Given a genus-$g$ Riemann surface with $n$ punctures and a
particular sewing of the surface from three-punctured spheres,
consider the generalized quiver gauge theory naturally associated to
it. Then, the AGT duality \cite{AGT} is as the following statement:
\emph{the conformal block for this sewing is the instanton part of
Nekrasov's partition function of this gauge theory \cite{Nekrasov1}.
Furthermore, the n-point function of the Liouville theory on this
Riemann surface is equal to the integral of the absolute value
squared of Nekrasov's full partition function of this gauge
theory.}\cite{AGT} The dictionary of this duality is in table 1.

\subsection{Surface and Loop Operators in $\mathcal{N}=2$ $SU(2)$ Gauge
Theories}

There are three gauge invariant operators in gauge theories which
can be obtained from compactifying of two M5 branes on Reimann
surfaces: surface, line and point operators
\cite{Gaiotto0911}\cite{AGGTV}. These operators are all connected to
M2 branes which are attached to M5 branes. For the present aim, we
will only review surface operators and loop operators.

The surface operators are defined by considering an M2 boundary
surface $\mathcal{S}$ to be embedded in the 4d space-time
$\mathbb{R}^4$ and localized as a point $z$ on $C$. As in
\cite{Gaiotto0911}\cite{AGGTV}, the expectation value of the
\emph{elementary} surface operator in the $N=2 \,SU(2)$ gauge theory
is related to an insertion of a degenerate primary operator
$V_{1,2}(z)=e^{(b/2)\phi(z)}$ into Liouville correlation function.
The notation of the degenerate primary operator will be clarified in
the next subsection. It should be reminded that we are now
considering gauge theory living on $\mathbb{R}^4$ which can be seen
as a ``chiral" part of the same gauge theory on $S^4$. By the
hemispherical stereographic projection of $S^4$ onto two copies of
$\mathbb{R}^4$, this scenario is shown clearly in Fig.3 of
\cite{AGGTV}.

The line or loop operators are represented by M2 brane boudaries that wrap a circle $\gamma$ on $C$, and extend along
an infinite line or closed loop $\mathcal{C}$ in $\mathbb{R}^4$. A Wilson-t'Hooft loop operator is labeled by the
circle $\gamma$ and can be computed in Liouville theory. It is proposed in \cite{AGGTV} that the expectation values of
loop operators are identical to associated monodromies in Liouville theory. The physical explanation is as following:
consider the annihilation of two identical surface operators in $S^4$, both are at the same position in $S^4$ and $C$,
except that one of them has traveled along a circle $\gamma$. Thus there exists a discontinuity between them. This
discontinuity defect can be identified as a monodromy in Liouville language. Now recall that the Nekrasov partition
function on $S^4$ is equal to the full Liouville correlation function, thus on $\mathbb{R}^4$, we are dealing with the
chiral conformal block of the Liouville correlation function. Finally, the monodromy can be recognized as the effect
that a chiral primary operator travels along a nontrivial circle $\gamma$ once.
\subsection{Loop operators in Liouville theory}
We will now review loop operators dual to those in 4d gauge theories. Modular bootstrap of Liouville theory will also
be briefly reviewed for further usage.
\subsubsection{Liouville Field Theory}
The action describing Liouville theory on an arbitrary genus $g$ Riemann surface with $n$ punctures $C_{g,n}$ is given
by:
\begin{equation}
S=\frac{1}{4\pi}\int d^2 z (g^{ab}\partial_a \phi \partial_b \phi + QR\phi + 4\pi \mu e^{2b\phi}),
\end{equation}
where $Q$ is the background charge, $\mu$ is the cosmological coupling constant. Liouville field theory(LFT) is a
conformal field theory if and only if $Q=b+1/b$ is satisfied. The central charge is: $c=1+6Q^2$ (See
reference~\cite{LFT} for a review of LFT). Primary fields are of the form $V_{\alpha}=e^{2\alpha \phi}$, and have
conformal weight
\begin{equation}
h_{\alpha}=\alpha(Q-\alpha).
\end{equation}
Note that primaries $V_{\alpha}$ and $V_{Q-\alpha}$ have the same conformal weight and are closely related. This will
bring some ambiguities in calculation of correlation functions. More precisely, the Liouville reflection amplitude
reads \cite{ZZ}
\begin{equation}
\mathcal{R}_L(\alpha)=-(\pi\mu\gamma(b^2))^{(Q-2\alpha)/b}\frac{\Gamma(1-(Q-2\alpha)b)
\Gamma(1-(\frac{Q-2\alpha}{b}))}{\Gamma(1+(Q-2\alpha)b)\Gamma(1+(\frac{Q-2\alpha}{b}))},
\end{equation}
and allows us to write $V_{\alpha}=\mathcal{R}_L(\alpha)V_{Q-\alpha}$, a relation which holds in any correlation
function. Here $\gamma(x)=\Gamma(x)/\Gamma(1-x)$.

Physical (unitary) representations are obtained for
\begin{equation}\label{conformal dim}
2\alpha = Q + i s,
\end{equation}
with $s\in \mathbb{R}$, which can be further restricted to $s\in
\mathbb{R}^+$ because of the reflection relation. They are the so
called non-degenerate spectrum (or continuum spectrum). There exist
degenerate representations which can be labeled by two coprime
numbers $(m,n)$. The charge (Liouville momentum) is given by
\begin{equation}
2\alpha=b^{-1}(1-m)+b(1-n).
\end{equation}
We denote the corresponding vertex operator at $z$ as $V_{m,n}(z)$. However, degenerate spectrums are not unitary
representations except the $(1,1)$ state which corresponds to the basic vacuum in LFT~\cite{Troost}~\cite{EST}.

The conformal bootstrap allows us to compute n-point correlation functions which can be formally written as:
\begin{equation}
\mathcal{Z}_{S^4}={\langle}\prod_{a=1}^{n}V_{m_a}(z_a){\rangle}_{C_{g,n}}=\int
d\nu(\alpha)\mathcal{\bar{F}}_{\alpha,E}^{(\sigma)}\mathcal{F}_{\alpha,E}^{(\sigma)},
\end{equation}
where $\mathcal{F}_{\alpha,E}^{(\sigma)}$ denotes conformal block associated to the sewing $\sigma$,
$\alpha\equiv\{\alpha_1,\dots,\alpha_{3g-3+n}\}$ label the internal Liouville momenta associated to the sewing of
conformal blocks, while $E\equiv\{m_1,\dots, m_n\}$ label the external Liouville momenta related to the masses of
hypermultiplets in $\mathcal{N}=2$ gauge theory. The measure $\nu(\alpha)$ includes for each trivalent graph dissection
of the conformal block. The explicit expression for $\nu(\alpha)$ was introduced in~\cite{Zamolodchikov:1995aa}, and
derived in~\cite{Teschner:Lecture}. After absorbing the prefactor into the conformal block, one will arrive at the
simple expression used in~\cite{AGGTV}:
\begin{equation}
\mathcal{Z}_{S^4}=\int d a_i |\langle\prod_{a=1}^{n}V_{m_a}(z_a)\rangle_{\{a_i\}}|^2,
\end{equation}
where the lower index $\{a_i\}$ labels the channel of conformal blocks and $2a_i = 2\alpha_i - Q$. This partition
function is non-chiral and invariant under the modular S transformation. However, as reviewed in last section, the loop
operator can be seen as an action of a chiral degenerate vertex operator on a conformal block of the full partition
function. Thus the main interesting thing is actually the block
\begin{equation}
\mathcal{Z}_{4d} = \langle\prod_{a=1}^{n}V_{m_a}(z_a)\rangle_{\{a_i\}},
\end{equation}
with an insertion of chiral degenerate vertex operator.

 For the convenience of computation, it is useful to introduce
another normalized conformal block $\mathcal{G}_{\alpha,
E}^{(\sigma)}$ which was first proposed by Ponsot and Teschner in
\cite{Teschner}. This conformal block can be related to
$\mathcal{F}_{\alpha,E}^{(\sigma)}$ by normalization, as in
\cite{DGOT}. The partition function in $S^4$ now can be written as:
\begin{equation}
\mathcal{\tilde{Z}}_{S^4} = \int d \mu(\alpha)\mathcal{\bar{G}}_{\alpha, E}^{(\sigma)}\mathcal{G}_{\alpha,
E}^{(\sigma)}.
\end{equation}
The partition function $\mathcal{\tilde{Z}}_{S^4}$ is slightly
different with the former one $\mathcal{Z}_{S^4}$ due to the
normalization, where the measure $d\mu(\alpha)$ is:
\begin{equation}
d \mu(\alpha)=\prod_{i=1}^{3g-3+n} d \alpha_i (-4 sin(2\pi a_i b)sin(2\pi a_i b^{-1})).
\end{equation}
Later on, we will see that this is just a modular S transformation
of the former up to an irrelevant normalization number. So conformal
blocks $\mathcal{Z}_{4d}$ and $\mathcal{G}_{\alpha, E}^{(\sigma)}$
are related by an modular $S$ transformation. This can be further
clarified via the 3d surgery method which we will focus on in
section 4.

\subsubsection{Loop Operators in Liouville Field Theory}
The computation strategy of loop operators in LFT is clear now.
First, introduce an identity operator in the conformal block.
Second, split it into two chiral degenerate operators $V_{1,2}(z)$
\footnote{The corresponding charged particles are in the spin 1/2
representation, in general case, the loop operator can be in
arbitrary spin representation of SU(2). For a $j/2$ spin loop
operator, the associate chiral degenerate operator is $V_{1,l}$ with
$l=2j+1$.}. Third, let one of these two operators round along a
circle which labels the loop one time. Finally, glue the two
operators back to identity. For Wilson line, this is very simple.
One will only consider $V_{1,2}(z)$ round another internal vertex
operator which can be associated to the sewing of conformal blocks
\footnote{Or equivalently, the thin tube which connects two pants
components of the Riemann surface $C$.} once~\cite{Sonoda:1988mf}.
For t'Hooft line, this becomes complicated because now $V_{1,2}(z)$
should travel around all the background, including all handles and
punctures. However, the computation of these quantities are all
considered and finished in~\cite{AGGTV} and also~\cite{DMO,DGOT}.
For details, one may refer to these two articles~\cite{DMO,DGOT}.
\subsubsection{Modular Bootstrap}
Another way to invoke the same LFT is the so called modular
bootstrap. This method was introduced by Zamolodchikov
brothers~\cite{ZZ} and further developed by Jego and
Troost~\cite{Troost} and Eguchi et al~\cite{EST,Eguchi}. The basic
ingredients are characters of highest weight states including
degenerate and non-degenerate ones. For non-degenerate
representations ($2\alpha = Q + is, s\in \mathbb{R}^+$), the
character and conformal dimension are
\begin{equation}
\chi_s(\tau)=\frac{q^{s^2/4}}{\eta(\tau)},\,\,\,\,\,\,\, h_s = \frac{1}{4}(Q^2 + s^2),
\end{equation}
where $\eta$ is the Dedekind function, $q = e^{i 2\pi \tau}$. For degenerate representations (which for nonrational $b$
have a single null vector at level $nm$), the character and conformal dimension are
\begin{align}
\chi_{m,n}(\tau) = \frac{q^{-(m/b +
nb)^2/4}-q^{-(m/b-nb)^2/4}}{\eta(\tau)},\nonumber\\
h_{m,n} = \frac{1}{4}(Q^2 - (m/b + nb)^2)\;.
\end{align}
The modular transformations of the characters are \cite{Troost, EST, Eguchi}
\begin{align}
\chi_s(-\frac{1}{\tau}) = \int_0^{\infty}S_{s}^{s'}\chi_{s'}(\tau) d
s', \,\,\,\,\,\,S^{\,\,\,\,s'}_{s} = \sqrt{2}cos(\pi s s'),\\
\chi_{m,n}(-\frac{1}{\tau}) = \int_{0}^{\infty}S_{m,n}^{s'} \chi_{s'}(\tau) d s',
\quad\quad S_{m,n}^{\,\,\,\,\,\,\,\,\,s'} = 2\sqrt{2}sinh(\pi m \frac{s'}{b})sinh(\pi n b s'),\\
S_{m,n}^{\,\,\,\,\,\,\,\,\,m',n'} = -2\sqrt{2}sin(\pi m b^{-1}(m'b^{-1}+n'b))sin(\pi n b (m'b^{-1}+n'b))\;.
\end{align}
The third modular S transformation for degenerate states can be
obtained from analytic continuation of the second one. However, as
we mentioned before, these degenerate representations are not
unitary. They become unitary only if there exists a bigger system in
which LFT as a subsystem. The most possible system is the
supersymmetric extension of LFT. Actually, it is well known
\cite{Eguchi} that supersymmetric Liouville theory does have unitary
degenerate representations. This implies that the good dual theory
for $\mathcal{N}=2$ SCFTs is a supersymmetric version of LFT. We
hope further studies will clarify this.

Notice that these modular S transformations come from the affine
algebra. It is just another representation of CFT since there exists
one to one correspondence between CFTs and quantum groups
representations \cite{Moore Seiberg}.
\section{From Chern-Simons to Liouville}
A Chern-Simons theory with compact gauge group on (2+1)d is an exact
dual description of (1+1)d WZW model with the same gauge group. This
is clarified in Witten's illuminating work \cite{CSW} two decades
ago. However, for non-compact group, this duality is far from clear
on both sides. Fortunately, for $SL(2,\mathbb{R})$ considered in
present situation, there are some important developments
\cite{CSW;nocompact, Carlip 91, Banados 94, Carlip 05} which we will
briefly review in the following text. The action of
$SL(2,\mathbb{R})$ Chern-Simons theory on 3d manifold $M$ can be
written as
\begin{equation}
I_{CS}[A] = \frac{k}{2\pi}\int_M {\rm Tr}\big\{ A\wedge d A + \frac{2}{3} A \wedge A \wedge A\big\},
\end{equation}
with $A$ takeing value in the Lie algebra space $\mathbb{H}$ of
$SL(2, \mathbb{R})$, $``\rm{Tr}"$ is an invariant form on
$\mathbb{H}$. If $M$ is compact, this action is gauge invariant, but
this is not the case when $M$ is a 3-manifold with boundary
$\Sigma$: under a gauge transformation
\begin{equation}
A = g^{-1} d g + g^{-1} \bar{A} g,
\end{equation}
the action $I_{CS}$ transforms as
\begin{equation}\label{Gauge trans of CS}
I_{CS}[A] = I_{CS}[\bar{A}]-\frac{k}{4\pi}\int_{\Sigma} {\rm Tr}((dg g^{-1})\wedge \bar{A}) - \frac{k}{12\pi}\int_M
{\rm Tr}(g^{-1}dg)^3 .
\end{equation}
Naively, if one chooses the 3-manifold as the topologically trivial
one: $\mathbb{R}\times D$, where $D$ is a disk, this means that
gauge potential $A$ in the bulk is a ``pure gauge"
\begin{equation}
A = g^{-1} d g .
\end{equation}
Substituting this into (\ref{Gauge trans of CS}), a straightforward
computation shows that the resulting action for $g$ is a WZW action
on the boundary $\partial M = \mathbb{R} \times S^1$ \cite{Fjelstad
Hwang}.

For a general 3-d manifold with boundary, it is necessary to
introduce proper boundary condition and boundary counterterm if one
would like to keep gauge invariance in the bulk. A proper boundary
condition should constrain half degrees of the phase space. However,
in Chern-Simons theory, it can be easily read off from the canonical
quantization that the gauge potentials $A$ are both canonical
positions and momenta. Thus the natural choice is to fix the value
of boundary gauge potential, or equivalently, choose a complex
structure on $\Sigma$. Without loss of generality, we fix the value
for $A_z$ and the boundary term now can be written as
\begin{equation}
I_{bdry}[A] = \frac{k}{4\pi} \int_{\Sigma} {\rm Tr} A_z A_{\bar{z}},
\end{equation}
which transforms as
\begin{equation}
I_{bdry}[A] = I_{bdry}[\bar{A}] + \frac{k}{4\pi}\int_{\Sigma} {\rm Tr}(\partial_z g g^{-1} \partial_{\bar{z}}g g^{-1} +
\partial_z g g^{-1}\bar{A}_{\bar{z}}+\partial_{\bar{z}} g g^{-1}\bar{A}_{z}).
\end{equation}
Now the full action transforms as
\begin{equation}
(I_{CS}+I_{bdry})[A] = (I_{CS}+I_{bdry})[\bar{A}] + k I^{+}_{WZW}[g^{-1}, \bar{A}],
\end{equation}
with the chiral WZW action
\begin{equation}
I^{+}_{WZW}[g^{-1}, \bar{A}] = \frac{1}{4\pi}\int_{\Sigma}{\rm Tr}(\partial_z g g^{-1} \partial_{\bar{z}}g g^{-1} -2
g^{-1}\partial_{\bar{z}} g\bar{A}_{z}) +\frac{1}{12\pi}\int_{M}{\rm Tr}(g^{-1}dg)^3.
\end{equation}
Note that now $g$ is a dynamical field on $\Sigma$, which implies
that pure gauge transformation in bulk becomes real symmetry on
boundary. The additional degree of freedom\footnote{Here, we refer
to the degrees of freedom of $g^{-1}$.} is just a result from
reducing the second class constraints to the first class
constraints.

To get a CFT dual description for Chern-Simons on 3-manifold $M$
without boundary, one can cut $M$ into two pieces $M_1$ and $M_2$
with the same boundary $\Sigma$; on each piece there is a
Chern-Simons. By gluing these two pieces back into $M$ carefully,
one can get a Chern-Simons on $M$. From the CFT side of view, one
just combines a chiral WZW theory and an anti-chiral WZW theory to a
non-chiral WZW theory on $\Sigma$. This is feasible when a CFT is
holomorphic factorizable \cite{Holomorphic Factorization}. The
``sewing" of WZW \cite{Sonoda:1988mf} models with gauge fields
$A_z^+$ and $A_{\bar{z}}^-$ can be realized using Hamiltonian
reduction method \cite{Hamiltonian Reduction, Ishibashi, Izawa}. In
the process of ``sewing", one should introduce additional
constraints. The simplest case is $A_z^{+}= A_{\bar{z}}^- = 0$
\footnote{One can treat this as an on-shell constraint, or the
classical constraint.}, say, the gauge fields vanish simultaneously.
Now the ``sewing" is trivial:
\begin{equation}
\partial_{-} J_{+} = 0, ~~~~\partial_{+} J_{-} = 0,
\end{equation}
where we have changed the labels $\partial_{+}\equiv \partial_{z},
\partial_{-}\equiv
\partial_{\bar{z}}$ for further simplicity. $J_{+}$ and $J_{-}$ areleft and
right Kac-Moody currents respectively
\begin{equation}
J_{+} = (\partial_{+}g)g^{-1}, ~~~~J_{-} = g^{-1}(\partial_{-}g).
\end{equation}
These are just equations of motion for left and right $SL(2, \mathbb{R})$ invariant vector fields $g^{-1}d g$ and $dg
g^{-1}$ respectively. Before rushing to the off-shell situation, we should keep under observation on gauge fields $A_+$
and $A_-$. These fields naturally introduce a complex structure and further define inner products on $M_1$ and $M_2$
\begin{equation}
\partial_{+}\mapsto\partial_{+} - A_{+},
~~~~\partial_{-}\mapsto\partial_{-} + A_{-},
\end{equation}
where the different definition comes from the opposite chirality.
 If one drags $A_{+}$ to $M_2$ (or drags $A_{-}$ to $M_1$), it is
necessary to change $A_{+}$ ($A_{-}$) to its right(left) invariant
form $g^{-1}A_{+}g$ ($g A_{-}g^{-1}$). Now we can introduce the
gauge-invariant action
\begin{eqnarray}
I[g, A_{+}, A_-] = I_{WZW}[g] &-&
\frac{k}{2\pi}\int_{\Sigma}\rm{Tr}\big\{(A_{-}(\partial_{+}g)g^{-1}-\sqrt{\mu})\nonumber\\&+&(g^{-1}\partial_{-}g
A^{+}-\sqrt{\mu})+A_{-}gA_{+}g^{-1}\big\},
\end{eqnarray}
where $\sqrt{\mu}$ is a constant valued in Cartan subalgebra
$\mathcal{H}$ of $SL(2,\mathbb{R})$, whose meaning will be clarified
in following text. The equations of motion are given by
\begin{align}
[D_{-}, D'_{+} - J_{+}] = 0,&&[D_{+}, D'_{-} + J_{-}] = 0, \\
J_{+} + gA_{+}g^{-1} - \sqrt{\mu}=0,&&J_{-} + gA_{-}g^{-1} - \sqrt{\mu}=0,
\end{align}
where $D_{-} = \partial_{-} + A_{-}$, $D_{+} = \partial_{+} + A_{+}$ and
\begin{equation}
D'_{+} = \partial_{+} - gA_{+}g^{-1}, ~~~~ D'_{-} = \partial_{-} + g^{-1} A_{-} g,
\end{equation}
reflecting the transition of the connection due to the ``sewing". If one identifies
\begin{equation}
A_{-} =  g A_{+} g^{-1}, ~~~~ A_{+} = g^{-1} A_{-} g,
\end{equation}
the first two equations of motion are just chiral anomaly equations:
\begin{equation}
[D_{-}, J_{+}] = F_{-+}, ~~~~ [D_{+}, J_{-}] = F_{+-}.
\end{equation}
 A good ``sewing" should be anomaly free, or equivalently, identifying two patches only
 up to a pure gauge transformation. So the gauge strength $F$ should
 vanish and the associated gauge connection is the flat connection.
 Because of this, one can at first set $A_{\pm} = 0$, then the theory will
 reduce to ordinary WZW action $I_{WZW}[g]$ with constraints:
 \begin{eqnarray}\label{constraints}
J_{+}  - \sqrt{\mu}=0,~~~~ J_{-} - \sqrt{\mu}=0.
 \end{eqnarray}
Now the derivation is straightforward, first let us parameterize $g\in SL(2, \mathbb{R})$ via the Gauss decomposition
\[ g = \left( \begin{array}{cc}
1 & v \\
0 & 1  \\
 \end{array} \right)\left( \begin{array}{cc}
e^{\phi} & 0\\
0 & e^{-\phi}  \\
 \end{array} \right)\left( \begin{array}{cc}
1 & 0 \\
\bar{v} & 1  \\
 \end{array} \right).\]
 In these coordinates, the action $I_{WZW}$ becomes
 \begin{equation}
 I = \frac{k}{8\pi}\int d z^2 \big\{ \partial_{z} \phi \partial_{\bar{z}}\phi + e^{-2\phi} \partial{\bar{z}v
 \partial_{z}\bar{v}}\big\}.
 \end{equation}
Second, substituting the constraints (\ref{constraints}) into the
action, one immediately obtain the Liouville action
\[I_{Liou} = I = \frac{k}{8\pi}\int d z^2 \big\{ \partial_{z} \phi \partial_{\bar{z}}\phi + \mu
e^{2\phi}\big\},\] where the meaning of $\mu$ is clear: though it is only an integral constant in WZW models, now it
has a physical meaning as the cosmological constant.

This connection between Chern-Simons and Liouville theory is rather rough. One should expect there is a dictionary from
Liouville to Chern-Simons. In Liouville side, given the knowledge of central charge, conformal dimensions of primary
fields, fusion rules and conformal blocks, one can totally determine the whole theory. Now the central subject is to
identify these objects in Chern-Simons theory.

\section{Links, Surgery and Wilson Loop}
In this section we define the monodromy of loop operator in
Liouville theory as a ratio of link invariants in Chern-Simons
theory which lives on $S^3$. We also consider t'Hooft loops in
$\mathcal{N}=4$ SYM theory.
\subsection{Assumptions}
A Wilson line could also be regarded as the space-time trajectory of
a charged particle, this point of view will immediately lift the
theory under consideration to 3d. Alternatively, one could realize
the effect of a loop operator in Liouville theory as a special knot
or link in 3d. Since each 3d manifold could be obtained by chopping
$S^3$ into pieces and then gluing them back after several
diffeomorphism~\cite{CSW}, it is sufficient to consider $S^3$ only.
However, it is important to bare in mind that Liouville theory is
living on the boundary of one piece of chopped 3d manifold. Tracking
the geometric diffeomorphism step by step, one could expect to
obtain the dual description of Liouville theory. This is exactly
what CSW-RCFT \cite{CSW} means. We expect this kind of chopping and
gluing back (surgery) should remain true in general situation even
for noncompact groups and irrational CFTs. Also, we expect that
there exist a one to one correspondence from Liouville conformal
blocks to quantum Hilbert spaces obtained by quantizing a three
dimensional theory. This is just a generalization of Witten's
statement on CSW-RCFT correspondence \cite{CSW}.

\subsection{Surgery}

The easiest way to construct a 3d manifold from 2d CFT is as
following: let the 2d Riemann surface $\Sigma$ with $n$ marked
points travel in 3d imaginary Euclidean space-time, then bend the
``time" direction to a circle, which will identify the original
surface and the final surface. The resulting manifold will be a
trivial bundle: $\Sigma\times S^1$ with $n$ Wilson lines living on
it. This process is shown in Fig.1.
\begin{figure}\label{fig0.eps}
\includegraphics[bb=20 437 588 790, width=15 cm, clip]{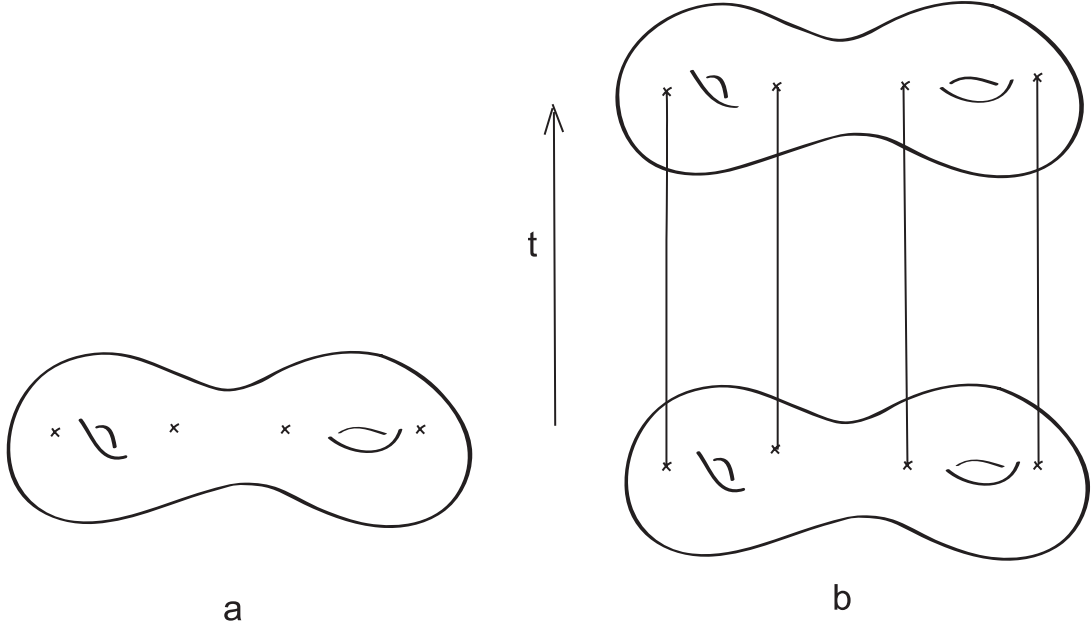}
 \caption{a) A
Riemann surface $\Sigma$ with four punctures. b) A segment of
$\Sigma \times S^1$ with four Wilson lines.}
\end{figure}
 In general, VEVs of Wilson
loops in this manifold are hard to calculate. However, Witten provided a very powerful method to deal with this problem
\cite{CSW}: the surgery method. This method admits to replace a 3d manifold with $n$ Wilson lines as $S^3$ with
different numbers of Wilson lines. Moreover, one could also ``erase'' all the $n$ Wilson lines by $n$ times sequent
surgeries, these operations will replace the original 3-manifold as a new manifold due to surgeries.

 For a concrete view of the surgery, consider a
very simple case\footnote{The process of this surgery is shown in
Fig.2, where $M$ stands for $S^2\times S^1$, $\tilde{M}$ stands for
$S^3$.}: $\Sigma$ is a Riemann sphere with no punctures. Now the
resulting manifold is an $S^2\times S^1$ , next draw a mathematical
loop $C$ on it and then cut the neighborhood $M_R $ of the loop, the
boundary of $M_R$ is simply a torus denoted as $T$. We pick a basis
of $H^1(T;\mathbb{Z})$ consisting of cycles a and b indicated in
Fig. 2b, then choose a diffeomorphism $S:~T\longrightarrow T$ as the
map $a\longrightarrow b, b\longrightarrow -a$. Then glue the changed
$M_R$ back to the remain $M_L$, one will immediately obtain an
$S^3$. This means that computations on $S^3$ are equivalent to
computations on $S^2\times S^1$ with a physical Wilson line where
the surgery was made. We will call this surgery of $S^3$ as
``modular surgery" hereafter. One can also recognize the special
diffeomorphism that exchanges a, b circles as nothing but the
modular $S$ transformation of the torus $T$. Now we want to explain
what this $S$ transformation means in 2d CFT. To do this, let us
look at the surgery more closely. It is easy to see that cutting the
neighborhood of $C$ corresponds to cutting $S^2$ into two disks $D$
and $D'$, then the partition function of the CFT on $S^2$ can be
obtained by gluing two CFTs (holomorphic CFT and anti-holomorphic
CFT, respectively) on different disks with a common boundary. Now
the meaning of the modular $S$ transformation is clear: it is just a
change of boundary condition of the CFT on $D'$. Recall that
partition function of a holomorphic factorizable CFT can always be
constructed by characters of the related affine algebra, the modular
$S$ transformation of the holomorphic character can be thought as
the consequence of the changing of boundary condition. Using this
surgery to get an $S^3$ corresponds to the conformal blocks used in
\cite{DGOT,Teschner}.
\begin{figure}\label{fig2}
\includegraphics[bb=21 254 583 762, width=15 cm, clip]{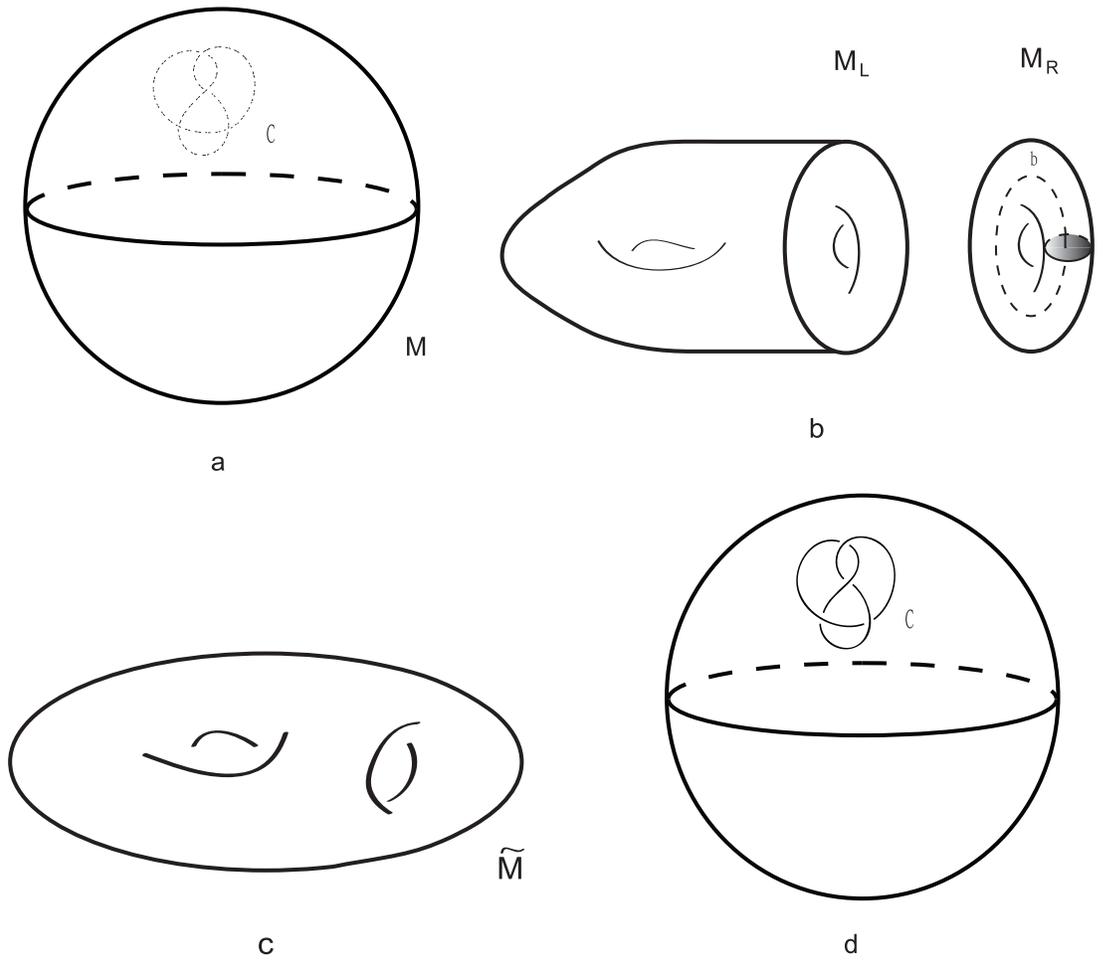}
\caption{In a), an imaginary loop $C$ was darwn on $M$. In b), the
neighborhood of $C$ was cut out, now the $M$ had been cut into $M_L$
and $M_R$ which is a solid torus. In c), the solid torus had been
glued back with $M_L$ after a diffeomorphism $K$, and formed a new
3-manifold $\tilde{M}$. In d), It follows that partition function in
$\tilde{M}$ can be obtained from that in $M$ with a Wilson loop $C$
and the knowledge of diffeomorphism $K$.}
\end{figure}
However, there is an easier way to obtain an $S^3$ from two
3-manifolds: picking two identical 3-balls $B_1, B_2$ with boundary
2-spheres $S^2_1, S^2_2$, gluing $S^2_1, S^2_2$ without
diffeomorphisms, one will immediately get an $S^3$. This corresponds
to the conformal blocks used in \cite{AGGTV}. We will call this
surgery of $S^3$ as ``simple surgery" hereafter. Now we recognize
the difference between these two kinds of blocks is nothing but a
modular transformation $S^{\,\,\,(-i2a)}_{1,1}$, the lower $(1,1)$
labels the basic vacuum, $is = 2 a$ can be read off immediately from
the expression of conformal dimension $h$ which shows in
\ref{conformal dim} and $2\alpha = Q + 2a$. For further convenience,
we do not distinguish the label $S^{\,\,\,\,\,\,(-i2a)}_{1,1}$ and
$S^{\,\,\,\,\,\,2a}_{1,1}$.

We now consider the standard surgery for generic 3d manifold
$M$(Fig.2a). To do so we first thicken $C$ to a ``tubular
neighborhood", a solid torus centered on $C$. Removing this solid
torus, $M$ is split into two pieces; the solid torus is called
$M_R$(Fig.2b), and the remainder is called $M_R$. One then makes a
diffeomorphism on the boundary of $M_R$ and glues $M_R$ and $M_L$
back together to get a new three manifold $\tilde{M}$(Fig.2c). The
canonical quantization of Chern-Simons theory on 3d manifold makes
it clear that Hilbert spaces $\mathscr{H}_L$ and $\mathscr{H}_R$,
canonically dual to one another, are associated with the boundaries
of $M_L$ and $M_R$. The path integrals on $M_L$ and $M_R$ give
vectors $\psi$ and $\chi$ in $\mathscr{H}_L$ and $\mathscr{H}_R$,
and the partition function on $M$ is just the natural pairing
$(\psi, \chi)$. If we act on the boundary $M_R$ with a
diffeomorphism $K$ before gluing $M_L$ and $M_R$ together, then
$\chi$ is replaced by $K\chi$ so $(\psi, \chi)$ is replaced by
$(\psi, K \chi)$. So the partition function \footnote{For more
details see the reference \cite{CSW}.} on $\tilde{M}$ can be related
to that on $M$(Fig.2d) as
\begin{equation}
Z(\tilde{M}) = \sum_j K_{0}^{\,\,\,j}\cdot Z(M; R_j),
\end{equation}
where $K_{0}^{\,\,\,j}$ is the diffeomorphism associated with the surgery. It is clear that if $M$ is $S^2\times S^1$
and $\tilde{M}$ is $S^3$, $K_{i}^{\,\,\,j}\equiv S_{i}^{\,\,\,j}$.

\subsection{Generalized Surgery}

Now we will consider the generalized surgery on 3d manifold $M$. In
this situation, before the surgery a Wilson line in the $R_i$
representation was already present on the imaginary circle $C$.
Surgery amounts to cutting out a neighborhood of $C$ and then gluing
it back in, and after this process the $R_i$ Wilson line will still
be present in $\tilde{M}$. Now the diffeomorphism also acts on the
$R_i$ Wilson line, then detailed analysis gives \cite{CSW}:
\begin{equation}\label{generalized surgery}
Z(\tilde{M}; R_j) = \sum_j K_{i}^{\,\,\,j}\cdot Z(M; R_j).
\end{equation}

\subsection{Path Integrals on $S^1 \times X$}

The three manifolds whose partition functions can be computed in a
particularly simple way, from the axioms of quantum field theory,
are those of the form $X\times S^1$, for various $X$. $X\times S^1$
can have a ``Hamiltonian" formalism if one treats the $S^1$ as the
``bended" time direction. This can be realized as following: one
constructs the Hilbert space $\mathscr{H}_X$ of $X$, then introduces
a ``time" direction represented by a unit interval $I = [0,1]$, and
then propagates the vector in $\mathscr{H}_X$ from ``time" 0 to
``time" 1. This operation is trivial, since the Chern-Simons theory
is a topological field theory. It has a vanishing Hamiltonian.
Finally, one obtains $X\times S^1$ by gluing $X\times\{0\}$ to
$X\times\{1\}$; this identifies the initial and final states, giving
a trace:
\begin{equation}\label{cylin}
Z(X\times S^1) = \rm{Tr}_{\mathscr{H}_X}(1) = \rm{dim} \mathscr{H}_X.
\end{equation}

Now we would like to review the dimension of $\mathscr{H}_{S^2 ,n}$,
the Riemann sphere with $n$ punctures. There are well-known results
for this question \cite{Verlinde, Moore Seiberg} and can be found
explicitly in ref. \cite{CSW}. We now copy these results as follows:

(I) For the Riemann sphere with no punctures (marked points), the Hilbert space is one dimensional.

(II) For the Riemann sphere with one puncture in a representation $R_i$, the Hilbert space is one dimensional only if
$R_i$ is trivial, and zero dimensional otherwise.

(III) For the Riemann sphere with two punctures with representation $R_i$ and $R_j$, the Hilbert space is one
dimensional if $R_j$ is the dual of $R_i$ and zero dimensional otherwise.

(IV) For the Riemann sphere with three punctures with representation $R_i$, $R_j$, and $R_k$, the dimension of
$\mathscr{H}_{S^2 ,3}$ is the Verlinde number $N_{ijk}$ .

(V) From the results of Verlinde \cite{Verlinde}, the dimension of
the physical Hilbert spaces for an arbitrary collection of punctures
on $S^2$ can be determined from a knowledge of the Verlinde number
$N_{ijk}$. This refers to the fusion rules of the CFT.

Using these results, one can immediately obtain the partition function on $S^2\times S^1$
\begin{equation}
Z(S^2\times S^1) = 1.
\end{equation}
However, we should note here that this result is obtained by a
normalization, the partition function of $S^2\times S^1$ can be
strictly obtained either by quantization of Chern-Simons theory
\cite{PI QCS} or the operator formalism \cite{Labastida 89a}.
Another thing that should be clarified here is that this partition
function is only the holomorphic part of the full partition
function, say, the holomorphic character. As in \cite{Labastida
89a}, the states of a basis of the Hilbert space are in one to one
correspondence with the characters of the CFT\footnote{There the
authors considered the cases for compact gauge groups, we assume
this is also true in the non-compact case. Actually, the derivation
for the characters of degenerate fields are quite parallel. However,
the generalization for non-degenerate fields still unclear and we
hope further works will clarify this. }. Moreover, if there exist
$N$ unknotted and unlinked Wilson lines on the given 3-manifold $M$,
then the wave function related to a surgery of $M$ is equivalent to
a conformal block of the $2N$ correlation functions up to a
normalization constant~\cite{Labastida 89b}. In this scenario, the
Wilson line operators also had been identified as the Verlinde
operators in the dual CFT since they satisfied the same fusion
algebra. A geometric description on this fusion algebra can be found
in \cite{Guadagnini}.

If we are given a diffeomorphism $K: X \rightarrow X$, then one can form the mapping cylinder $X\times_K S^1$ by
identifying $x\times\{1\}$ with $K(x) \times \{1\}$ for every $x\in X$. The initial and final states are identified via
$K$, so the generalization of (\ref{cylin}) is
\begin{equation}
Z(X\times_K S^1) = {\rm Tr}_{\mathscr{H}_X}(K).
\end{equation}

For $X$ is $S^2$ with some punctures $P_a, a=1\dots s$ to which
representations $R_{i(a)}$ are assigned, we can use the above
results for Riemann sphere with punctures. We denote the Hilbert
space as $\mathscr{H}_{S^2;<R>}$ for $<R>$ representing the
collection of the punctures with representations. The partition
function is:
\begin{equation}
Z(S^2\times S^1; <R>) = {\rm dim} \mathscr{H}_{S^2; <R>}.
\end{equation}
Then for one puncture with representation $R_a$
\begin{equation}
Z(S^2\times S^1; R_a) = \delta_{a,0}.
\end{equation}
For two punctures with representation $R_a$ and $R_b$
\begin{equation}
Z(S^2\times S^1; R_a , R_b) = g_{ab},
\end{equation}
where $g_{ab}$ is defined as 1 if $R_b$ is the dual of $R_a$ and 0
otherwise. For three punctures with representation $R_a$, $R_b$ and
$R_c$
\begin{equation}
Z(S^2\times S^1; R_a , R_b, R_c) = N_{abc}.
\end{equation}
The Verlinde number $N_{abc}$ can be obtained either from loop
algebra of unknotted and unlinked loops in 3d~\cite{Labastida 89a}
or from the fusion algebra in 2d CFT.

\subsection{Hopf Links and Wilson Loop}
Now we can examine the above things on $S^3$. Using the surgery from
$S^2\times S^1$ to $S^3$, one can easily get
\begin{align}
Z(S^3) = \sum_j S_{0}^{\,\,\,j}Z(S^2\times S^1; R_j) = \sum_j
S_{0}^{\,\,\,j} \delta_{j,0} = S_{0,\,0}, \\
Z(S^3; R_j) = \sum_i S_{0}^{\,\,\,i}Z(S^2\times S^1; R_i, R_j) =
\sum_i S_{0}^{\,\,\,i} g_{ij} = S_{0,\,j},\\
Z(S^3, R_j, R_k)=\sum_i S_{0}^{\,\,\,i}Z(S^2\times S^1; R_i, R_j, R_k) = \sum_i S_{0}^{\,\,\,i} N_{ijk}.
\end{align}
The left-hand side of the last equation can be independently
calculated by cutting and gluing as in ref~\cite{CSW}, and the
result is
\begin{equation}
Z(S^3, R_j, R_k) = \frac{Z(S^3; R_j) Z(S^3; R_k)}{Z(S^3)} = \frac{S_{0,\,j}S_{0,\,k}}{S_{0,\,0}}.
\end{equation}
This is a special case of Verlinde formalism,
\begin{equation}
\frac{S_{0,\,j}S_{0,\,k}}{S_{0,\,0}} = \sum_i S_{0}^{\,\,\,i} N_{ijk}.
\end{equation}
 It has an obvious
meaning that two knots have been fused to a single knot. One can
recognize this as the basic fusion rule for the CFT. As a simple
check, we consider the fusion of one degenerate state $V_{1,2}$ and
one non-degenerate state $V_{\alpha}$. In Chern-Simons theory, this
can be identified as the action of taking one unknotted loop to be
close to another one with associated representations. The result is
simply what we just obtained
\begin{align*}
\frac{S_{1,1}^{~~1,2}S_{1,1}^{~~2a}}{S_{1,1}^{~~1,1}} & = -2cos(\pi b^2)(-2\sqrt{2}sin(\pi b^{-1} 2a)sin(\pi b 2a))
\\&= -2\sqrt{2}sin(\pi b^{-1} (2a+b))sin(\pi b (2a+b))\\&-2\sqrt{2}sin(\pi b^{-1} (2a-b))sin(\pi b (2a-b)) \\&=
S_{1,1}^{~~2a+b}+S_{1,1}^{~~2a-b},
\end{align*}
which is the fusion rule
\[
[V_{1,2}]\times[V_{\alpha}] = [V_{\alpha + \frac{b}{2}}]+[V_{\alpha - \frac{b}{2}}],
\]
where $[V_{\alpha}]$ denotes the whole Verma module for the primary
field $V_{\alpha}$. More general fusion rules can be derived
straightforwardly.
\begin{figure}\label{fig3}
\includegraphics[bb=33 527 555 772, width=15 cm, clip]{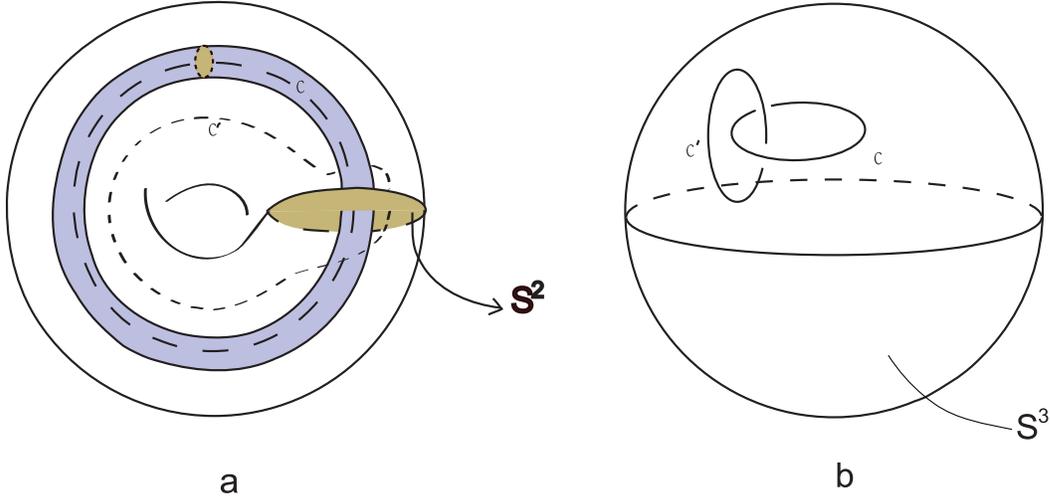}
\caption{a) In $S^2\times S^1$, two loops $C$ and $C'$ rounded the
uncontractable circle and braided with each other, the generalized
surgery was taken on $C$. b) After surgery, $C$ and $C'$ become a
linked Hopf link in $S^3$.}
\end{figure}
 One can also
generalize this to the generalized surgery situations. Consider the
case that there exist two braided loops in $S^2\times S^1$ in
representations $R_a$ and $R_b$ as showing in Fig.3a, making the
generalized surgery on $R_b$ circle and one gets a Hopf link $L(R_a,
R_b)$ on $S^3$ as shown in Fig.3b. The usage of the formula for
generalized surgery (\ref{generalized surgery}) therefore determines
the partition function of $S^3$ with a pair of linked Wilson lines:
\begin{equation}\label{Hopf link}
Z(S^3; L(R_i;R_j)) = \sum_k S^{\,\,\,k}_{i} Z(S^2\times S^1; R_k, R_j) = S_{i,\, j}.
\end{equation}
Again, one can easily obtain the entire fusion algebra by gluing two Hopf links to a satellite(or connected sum)
link~\cite{CSW, Guadagnini}
\begin{equation}
\frac{S_{i,\,j}S_{i,\,k}}{S_{0,\,i}} = \sum_l S_{i}^{\,\,\,l} N_{ljk}.
\end{equation}

 Since the Hopf link corresponds to the Wilson loop
operator in LFT, we now compute it explicitly using the modular S
transformation of LFT. The holonomy of the Wilson loop associated
with representation 1/2 is the phase factor of taking $V_{1,2}$
around the internal vertex operator $V_{\alpha}$ exact once. This
can be considered as the ratio of the final conformal block due to
the operation and the initial conformal block. This process can be
lifted to $S^2\times S^1$ as two braided Wilson lines. So the
holonomy can be obtained by comparing the ``initial" partition
function (without Wilson loop) with the ``final" one (with Wilson
loop). The ``final'' partition function corresponds to a Hopf link
invariant in $S^3$
\begin{equation}
S_{1,2}^{\,\,\,\,\,\,2a} = -2\sqrt{2} sin(2\pi b^{-1} a) sin(4\pi b a).
\end{equation}
It has the meaning of partition function only if one has normalized the vacuum partition function on $S^2\times S^1$ to
unity.  It is easy to write down the ``initial'' partition function (two unknotted loops)
\begin{equation}
S_{1,1}^{\,\,\,\,\,\,2a}\frac{S_{1,1}^{\,\,\,\,\,\,1,2}}{S_{1,1}^{\,\,\,\,\,\,1,1}} = -4\sqrt{2} sin(2\pi b^{-1} a)
sin(2\pi b a)cos(\pi b Q),
\end{equation}
where
\begin{equation}\label{Normalization}
\frac{S_{1,1}^{\,\,\,\,\,\,1,1}}{S_{1,2}^{\,\,\,\,\,\,1,1}} = \frac{1}{2cos(\pi b Q)},
\end{equation}
is the quantum dimension of chiral degenerate vertex operator
$V_{1,2}$.  Now one arrives at the holonomy $h_{1,2;\,\alpha}$ of
$V_{1,2}$ rounding a nontrivial loop
\begin{equation}
h_{1,2;\,\alpha} =
\frac{S_{1,2}^{\,\,\,\,\,\,2a}}{S_{1,1}^{\,\,\,\,\,\,2a}}\frac{S_{1,1}^{\,\,\,\,\,\,1,1}}{S_{1,2}^{\,\,\,\,\,\,1,1}} =
\frac{cos(2\pi b a)}{cos(\pi b Q)}.
\end{equation}
This is exactly the result obtained in \cite{DGOT, Pestun}. Now the
generalization to spin $j$ particle is straightforward, as in
\cite{DGOT}. One can replace the degenerate operator by
$V_{1,2j+1}$, then the associated monodromy can be obtained from
modular S transformations as follows:
\begin{equation}
h_{1,2j+1;\, \alpha} =
\frac{S_{1,2j+1}^{\,\,\,\,\,\,~~~~2a}}{S_{1,1}^{\,\,\,\,\,\,2a}}\frac{S_{1,1}^{\,\,\,\,\,\,1,1}}{S_{1,2j+1}^{\,\,\,\,\,\,~~~~1,1}}
= \frac{sin(2\pi b(2j+1) a)}{sin(2\pi b a)}\frac{sin(\pi b Q)}{sin(\pi (2j+1) b Q)}.
\end{equation}
Again, this agrees with the result in \cite{DGOT, Pestun}. We now
claim that this calculation is valid for all Riemann surfaces. The
reason is that the Wilson loop rounds only on one tube of the Rieman
surface, and one can cut the Riemann surface to its pants
decompositions, draw a Wilson loop on the specified pants and glue
it back. This process should not be interfered with other parts of
the Riemann surface, so it is sufficient to compute the Wilson loop
on Riemann sphere.
\begin{figure}
\includegraphics[bb=57 284 550 520, width=15 cm, clip]{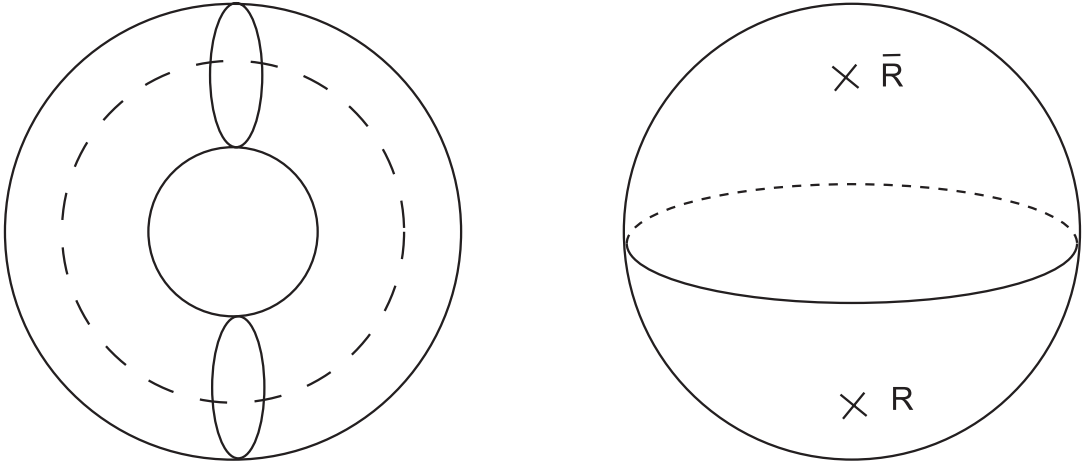}
\caption{A torus can be cut into two identical spheres with two
punctures which are attached with conjugated representations $R$ and
$\bar{R}$.}
\end{figure}

\begin{figure}
\includegraphics[bb=40 293 573 777, width=15 cm, clip]{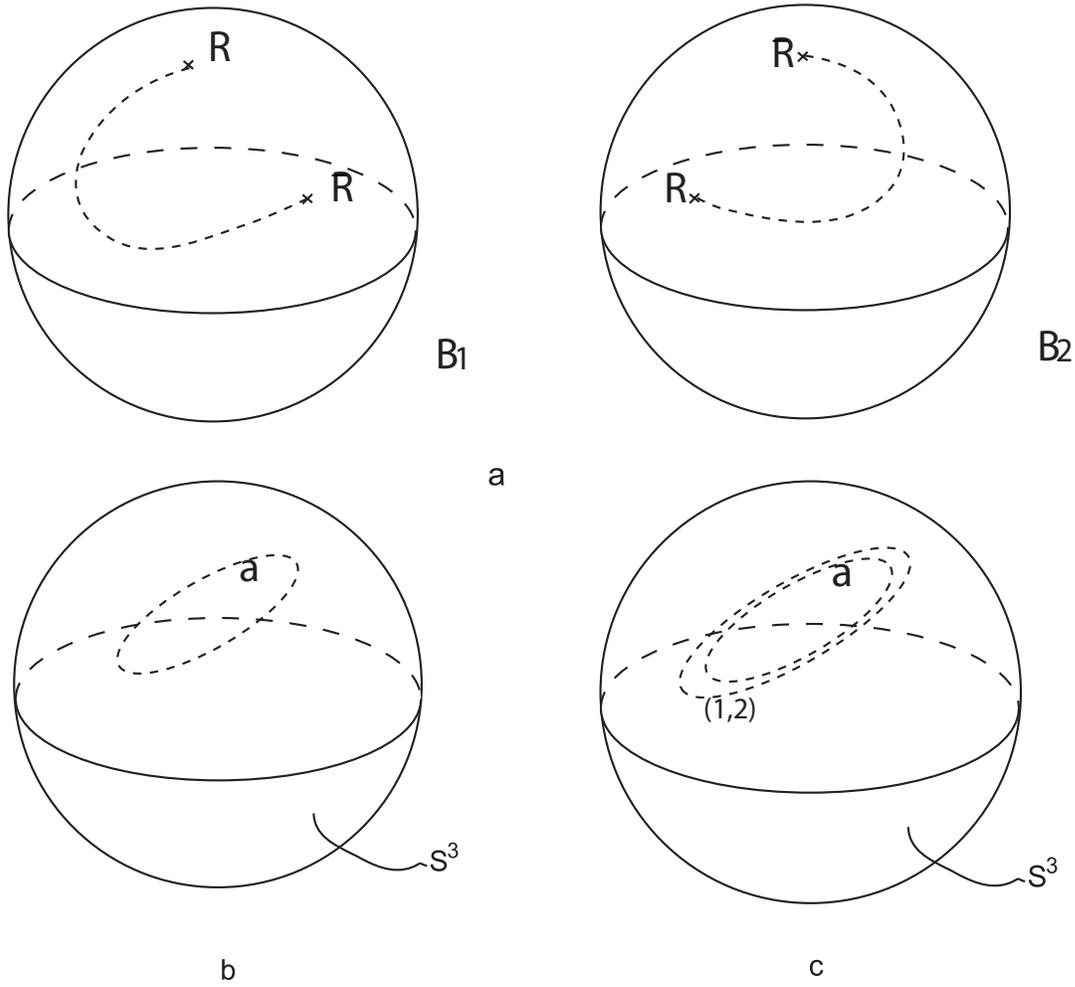}
\caption{a) The boundary of two balls $B_1$ and $B_2$ are identified
with spheres with two conjugated punctures, these two punctures are
endpoints of a Wilson lines in the bulk. b) By gluing $B_1$ and
$B_2$ back into $S^3$, one gets an $S^3$ with a single Wilson
loop.\,\,\,\, c) t'Hooft loop generated by (1,2) now can be seen as
the (1,2) Wilson loop surrounding the $a$ loop.}
\end{figure}

\subsection{t'Hooft loop in $\mathcal{N}=4$ SYM}
So far we have considered the contribution of Wilson loop. The
topology of the Riemann surface will be highly involved in the
computation of t'Hooft loops or Dyonic loops. We now consider the
simplest case: t'Hooft loop in $\mathcal{N}=4$ super Yang-Mills
theory (SYM).

The topology of the associated Riemann surface for $\mathcal{N}=4$
is a torus. For a torus, one could not obtain a simple 3-manifold by
rotating around a circle since $T^2\times S^1$ is a little hard to
deal with. The situation becomes more serious if one considers
higher genus geometry. However, there is a simple operation in CFT,
the sewing operation~\cite{Sonoda:1988mf}. A torus CFT can be sewed
from the same CFT liveing on a two-punctured sphere. The correlation
function on torus can also be obtained from the sphere CFT by the
sewing procedure. The translation from Liouville to Chern-Simons for
this sewing procedure is simple as we will clarify below. Cut a
torus into two identical cylinders with the same topology of
two-punctured spheres $S_{1}, S_{2}$ as in Fig.4. These two
punctures have been added representations conjugate to each other.
Instead of using the ``modular surgery'' method, we now prefer to
the ``simple surgery'' method, say, we identify $S_{1}$ and $S_{2}$
as boundaries of two identical balls $B_1$ and $B_2$, then in each
ball there is a half Wilson loop with two end points as the
punctures on the boundary (Fig.5a). Finally, glue $B_1$ and $B_2$ to
$S^3$ with a Wilson loop, as in Fig.5b. The result is really
nontrivial since the topology of torus (the genus) now becomes a
Wilson loop on $S^{3}$.

It may be a little subtle that we have used the ``simple surgery''.
However, one can also obtain the same result by the ``modular
surgery''. Let us show how this can be done. First, cut the torus
into two pieces as in Fig. 4. Instead of treating a cylinder as
two-punctured sphere, we let the cylinders deform to two
one-punctured disks\footnote{These two punctures are attached with
conjugated representations.}. Second, rounding each punctured disk
to form a solid torus within which a Wilson loop rounds the $b$
circle of the solid torus. Third, making the modular $S$
transformation for one solid torus (leaving the other invariant),
thus the Wilson loop within it now rounds the $a$ circle of the
solid torus. Finally, gluing both tori, one can get an $S^3$ with a
single Wilson loop since one should join two Wilson loops together
following the spirit of surgery. This reconstructs the result we
obtained using the ``simple surgery''.

Now the t'Hooft loop can be lifted to $S^3$ easily. Since $V_{1,2}$
should round all the background once, which is equivalent to drawing
a parallel Wilson loop of the one corresponding with the torus
geometry (Fig.5c). If one denotes the holomorphic conformal block
for the torus LFT without t'Hooft loop as $\mathcal{Z}(a)$ , where
$a$ denotes the representation of the intermediate state of the
sewing, then the monodromy of the t'Hooft loop generated by chiral
degenerate operator $V_{1,2}$ can also be computed by the sewing
procedure or directly by using fusion and braiding moves as in
\cite{AGGTV, DGOT}.  The chiral degenerate operator will only affect
the holomorphic part of partition functions which have a natural
representation as a partition function of knots or links invariant
on $S^3$, thus the holomorphic partition function $\mathcal{Z}(a)$
should be given as
\begin{equation}
\mathcal{Z}(a) = S_{1,1}^{~~2a},
\end{equation}
up to a normalization factor. The action of the t'Hooft loop will be given by the fusion of two loops
\begin{align}
\frac{S_{1,1}^{~~1,2}S_{1,1}^{~~2a}}{S_{1,1}^{~~1,1}} = S_{1,1}^{~~2a+b}+S_{1,1}^{~~2a-b} =
\mathcal{Z}(a+b/2)+\mathcal{Z}(a-b/2).
\end{align}
However, the contribution from ``zero mode'' $V_{\alpha} = V_{1,1}$
should be normalized, this gives the same factor
$N=\frac{1}{2cos(\pi b Q)}$, as in (\ref{Normalization}). The result
also matches with that in \cite{AGGTV, DGOT}. The generalization to
high genus topology is straightforward: each genus gives a
representation-attached circle on $S^3$, fusing these circles to the
degenerate one. The fusion rules will give the correct contribution
of the t'Hooft loop.
\section{Conclusions and Discussion}
In this note we have considered the three dimensional $SL(2,\mathbb{R})$ Chern-Simons description for Liouville field
theory which duals to an $\mathcal{N}=2$ SCFT in four dimensions by AGT conjecture. We have pointed out equivalence
between Chern-Simons and LFT from several points of view: actions, Hilbert spaces, conformal blocks, loop algebra and
fusion rules. Using these equivalent relations, we computed the contribution of Wilson loop in general case, also, we
give a simplest check on t'Hooft loop for $\mathcal{N} = 4$ SYM, the results we obtained match with those in
\cite{AGGTV, DGOT}.

The generalization of this Chern-Simons description of toda field theory just needs a change of gauge group, for
$A_{N-1}$ toda theory which duals to $SU(N)$ quiver gauge theories in 4d, the gauge group is $SL(N,\mathbb{R})$ for
Chern-Simons theory. Now loop operators in $SU(N)$ quiver gauge theories can be computed in Chern-Simons if given the
modular properties of the affine algebra of $SL(N,\mathbb{R})$. We are now preparing for this work.

The Chern-Simons/Liouville duality itself is far from a completed
one. There are many problems to be resolved. The first emergent
problem is how to derive DOZZ~\cite{Dorn:1994xn,ZZ} formula in
Chern-Simons theory, since this is the building block of the
Liouville theory. In principle, this can be done by canonical
quantization or using the operator formalism of Chern-Simons theory,
but it still needs a concrete work.

Second, does this Chern-Simons theory have an specified physical
origin instead of being a tool for calculations? One guess is there
may exist an origin, the M2 branes. This is a natural guess since we
are dealing with the systems which come from the configuration of M2
and M5 branes. Actually, Ooguri and Vafa had considered a similar
configuration a decade ago~\cite{Ooguri Vafa}. Another evidence is
that the Chern-Simons theory should be supersymmetric extended since
the corresponding LFT should be supersymmetric, in order to cure the
non-unitarity of degenerate representations which we used in the
context. Thus the Chern-Simons theory can be supersymmetric, which
could be related to the ABJM or BLG description~\cite{BLG, ABJM} for
M2 branes.

Third, if one can extract Chern-Simons theory from LFT, then following the spirit of the AGT duality, there should be a
direct path from $\mathcal{N}=2$ SCFTs to the same Chern-Simons theory. Thus the Chern-Simons theory plays the role of
a bridge which connects LFT with $\mathcal{N}=2$ SCFTs. Furthermore, if this connection is found, it will strongly
imply the duality between M2 and M5 branes in general construction.

Of course, there are other problems, for example, the relations between Chern-Simons and topological strings, also and
matrix theories, which still need to be clarified. S-duality in Liouville or SCFTs also should have a cousin in
Chern-Simons theory, which may be closely related to mirror symmetry in three dimension. We hope future works will
clarify these problems.

\section*{Acknowledgement}
We would like to thank Yan Liu for collaborations at the initial
stage of this project. We are particularly grateful to Prof. Miao Li
and Prof. Ming Yu for warmhearted support.
\renewcommand{\baselinestretch}{1}
\small\normalsize
\bibliography{All}{}

\providecommand{\href}[2]{#2}\begingroup\raggedright\begin{thebibliography}{10}

\bibitem{CSW}
E.~Witten, ``{Quantum Field Theory and Jones Polynomial},''
  \href{http://dx.doi.org/10.1007/BF01238857}{{\em Commun. Math. Phys.} {\bf
  121(3)} (1989), 351-399}

\bibitem{Gaiotto}
D.~Gaiotto, ``{$\mathcal{N}=2$ dualities}, ''
\href{http://arxiv.org/abs/0904.2715}{{\tt arXiv:0904.2715 [hep-th]}}.

\bibitem{Gaiotto0911}
D.~Gaiotto, ``{Surface Operators in N=2 4d Gauge Theories}, ''
\href{http://arxiv.org/abs/0911.1316}{{\tt arXiv:0911.1316 [hep-th]}}.

\bibitem{GaMa}
D.~Gaiotto, J.~Maldacena ``{The gravity duals of $\mathcal{N}=2$ superconformal field theories,} ''
\href{http://arxiv.org/abs/0904.4466}{{\tt arXiv:0904.4466 [hep-th]}}.

\bibitem{AGT}
L.~F.~Alday, D.~Gaiotto, Y.~Tachikawa, ``{Liouville Correlation Functions from Four-dimensional Gauge Theories},''
\href{http://arxiv.org/abs/0906.3219}{{\tt arXiv:0906.3219 [hep-th]}}.

\bibitem{AGGTV}
L.~F.~Alday, D.~Gaiotto, S.~Gukov, Y.~Tachikawa, H.~Verlinde, ``{Loop and surface operators in $\mathcal{N}=2$ gauge theory and Liouville modular geometry},''
\href{http://arxiv.org/abs/0909.0945}{{\tt arXiv:0909.0945 [hep-th]}}.

\bibitem{DMO}
N.~Drukker, D. R.~ Morrison and T.~Okuda, ``{Loop operators and S-duality from curves on Riemann surfaces},''
\href{http://arxiv.org/abs/0907.2593}{{\tt arXiv:0907.2593 [hep-th]}}.

\bibitem{DGOT}
N.~Drukker, J.~Gomis, T.~Okuda, and J.~Teschner, ``{Gauge Theory Loop Operators and Liouville Theory},''
\href{http://arxiv.org/abs/0909.1105}{{\tt arXiv:0909.1105 [hep-th]}}.

\bibitem{DV}
R.~Dijkgraaf and C.~Vafa, ``{Toda Theories, Matrix Models, Topological Strings, and $\mathcal{N}=2$ Gauge Systems},''
\href{http://arxiv.org/abs/0909.2453}{{\tt arXiv:0909.2453 [hep-th]}}.

\bibitem{Wyllard}
N.~Wyllard, ``{$A_{N-1}$ conformal Toda field theory correlation functions from conformal $\mathcal{N}$=2 $SU(N)$ quiver gauge theories},''
\href{http://arxiv.org/abs/0907.2189}{{\tt arXiv:0907.2189 [hep-th]}}.

\bibitem{Nanopoulos}
D.~Nanopoulos and D.~Xie, ``{N=2 SU Quiver with USP Ends or SU Ends with Antisymmetric Matter},''
\href{http://dx.doi.org/10.1088/1126-6708/2009/08/108}{{\em JHEP} {\bf 0908:108,2009.}}
\href{http://arxiv.org/abs/0907.1651}{{\tt arXiv:0907.1651 [hep-th]}}.
D.~Nanopoulos and D.~Xie, ``{On Crossing Symmmetry and Modular Invariance in Conformal Field Theory and S Duality in Gauge Theory},''
\href{http://arxiv.org/abs/0908.4409}{{\tt arXiv:0908.4409 [hep-th]}}.

\bibitem{Mironov etal}
A.~Marshakov, A.~Mironov, A.~Morozov, ``{On Combinatorial Expansions of Conformal Blocks},''
\href{http://arxiv.org/abs/0907.3946}{{\tt arXiv:0907.3946 [hep-th]}}.
A.~Mironov, S.~Mironov, A.~Morozov, A.~Morozov, ``{CFT exercises for the needs of AGT},''
\href{http://arxiv.org/abs/0908.2064}{{\tt arXiv:0908.2064 [hep-th]}}.
A.~Mironov, A.~Morozov, ``{The Power of Nekrasov Functions},''
\href{http://arxiv.org/abs/0908.2190}{{\tt arXiv:0908.2190 [hep-th]}}.
A.~Mironov, A.~Morozov, ``{On AGT relation in the case of U(3)},''
\href{http://arxiv.org/abs/0908.2569}{{\tt arXiv:0908.2569 [hep-th]}}.
A.~Marshakov, A.~Mironov, A.~Morozov, ``{On non-conformal limit of the AGT relations},''
\href{http://arxiv.org/abs/0909.2052}{{\tt arXiv:0909.2052 [hep-th]}}.
A.~Marshakov, A.~Mironov, A.~Morozov, ``{Zamolodchikov asymptotic formula and instanton expansion in N=2 SUSY $N_f=2N_c$ QCD},''
\href{http://arxiv.org/abs/0909.3338}{{\tt arXiv:0909.3338 [hep-th]}}.
A.~Mironov and A.~Morozov, ``{ Proving AGT relations in the large-c limit},''
\href{http://arxiv.org/abs/0909.3531}{{\tt arXiv:0909.3531 [hep-th]}}.
A.~Mironov and A.~Morozov, ``{ Nekrasov Functions and Exact Bohr-Zommerfeld Integrals},''
\href{http://arxiv.org/abs/0910.5670}{{\tt arXiv:0910.5670 [hep-th]}}.
V.~Alba and A.~Morozov, ``{Non-conformal limit of AGT relation from the 1-point torus conformal block},''
\href{http://arxiv.org/abs/0911.0363}{{\tt arXiv:0911.0363 [hep-th]}}.

\bibitem{Benini1}
F.~ Benini, S.~ Benvenuti and Y.~ Tachikawa,``{Webs of five-branes and N=2 superconformal field theories.
},''
\href{http://arxiv.org/abs/0906.0359}{{\tt arXiv:0906.0359 [hep-th]}}.

\bibitem{Benini2}
F.~ Benini, Y.~ Tachikawa and B.~Wecht,``{Sicilian gauge theories and N=1 dualities.
},''
\href{http://arxiv.org/abs/0909.1327}{{\tt arXiv:0909.1327 [hep-th]}}.

\bibitem{Maruyoshi}
K.~Maruyoshi, M.~Taki, S.~Terashima, F.~Yagi, ``{New Seiberg Dualities from N=2 Dualities},''
\href{http://dx.doi.org/10.1088/1126-6708/2009/09/086}{{\em JHEP} {\bf 0909:031,2009}}
\href{http://arxiv.org/abs/0907.2625}{{\tt arXiv:0907.2625 [hep-th]}}.

\bibitem{Gaiotto 0908}
D.~Gaiotto, ``{Asymptotically free N=2 theories and irregular conformal blocks},''
\href{http://arxiv.org/abs/0908.0307}{{\tt arXiv:0908.0307 [hep-th]}}.

\bibitem{Iguri Nunez}
S.~M.~Iguri, C.~A.~Nunez, ``{Coulomb integrals and conformal blocks in the AdS3-WZNW model},''
\href{http://arxiv.org/abs/0908.3460}{{\tt arXiv:0908.3460 [hep-th]}}.

\bibitem{Nekrasov Shatashvili}
N.~A.~Nekrasov and S.~L~Shatashvili, ``{Quantization of Integrable Systems and Four Dimensional Gauge Theories},''
\href{http://arxiv.org/abs/0908.4052}{{\tt arXiv:0908.4052 [hep-th]}}.

\bibitem{Jorjadze}
G.~Jorjadze, ``{Singular Liouville fields and spiky strings in $\mathbb{R}^{1,2}$ and $SL(2,\mathbb{R})$},''
\href{http://arxiv.org/abs/0909.0350}{{\tt arXiv:0909.0350 [hep-th]}}.

\bibitem{Poghossian}
R.~Poghossian, ``{Recursion relations in CFT and N=2 SYM theory},''
\href{http://arxiv.org/abs/0909.3412}{{\tt arXiv:0909.3412 [hep-th]}}.

\bibitem{Bonelli}
G.~Bonelli, A.~Tanzini, ``{Hitchin systems, N=2 gauge theories and W-gravity},''
\href{http://arxiv.org/abs/0909.4031}{{\tt arXiv:0909.4031 [hep-th]}}.

\bibitem{Giombi Pestun}
S.~Giombi, V.~Pestun, ``{The 1/2 BPS 't Hooft loops in N=4 SYM as instantons in 2d Yang-Mills},''
\href{http://arxiv.org/abs/0909.4272}{{\tt arXiv:0909.4272 [hep-th]}}.

\bibitem{Alday}
L.~Alday, F.~Benini, Y,~Tachikawa, ``{Liouville/Toda central charges from M5-branes},''
\href{http://arxiv.org/abs/0909.4776}{{\tt arXiv:0909.4776 [hep-th]}}.

\bibitem{Gadde}
A.~Gadde, E.~Pomoni, L.~Rastelli, S.~Razamat, ``{S-duality and 2d Topological QFT.},''
\href{http://arxiv.org/abs/0910.2225}{{\tt arXiv:0910.2225 [hep-th]}}.

\bibitem{Yang Zhou}
Y.~Zhou, ``{A Note on Wilson Loop in N=2 Quiver/M theory Gravity Duality.},''
\href{http://arxiv.org/abs/0910.4234}{{\tt arXiv:0910.4234 [hep-th]}}.

\bibitem{Yamada}
H.~Awata and Y.~Yamada, ``{Five-dimensional AGT Conjecture and the Deformed Virasoro Algebra},''
\href{http://arxiv.org/abs/0910.4431}{{\tt arXiv:0910.4431 [hep-th]}}.

\bibitem{Papadodimas}
K.~Papadodimas, ``{Topological Anti-Topological Fusion in Four-Dimensional Superconformal Field Theories},''
\href{http://arxiv.org/abs/0910.4963}{{\tt arXiv:0910.4963 [hep-th]}}.

\bibitem{Guadagnini}
E.~Guadagnini, ``{The Universal Link Polynomial},''
\href{http://dx.doi.org/10.1142/S0217751X92000417}{{\em International Journal of Modern Physics} {\bf A7 5}
  (1992)  877--945}
  
\bibitem{LFT}
Y.~Nakayama, ``{Liouville field theory: A decade after the revolution},''
  \href{http://dx.doi.org/10.1142/S0217751X04019500}{{\em Int. J. Mod. Phys.}
  {\bf A19} (2004)  2771--2930},
\href{http://arxiv.org/abs/hep-th/0402009}{{\tt arXiv:hep-th/0402009}}.

\bibitem{Dorn:1994xn}
H.~Dorn and H.-J. Otto, ``{Two and three-point functions in Liouville
  theory},'' \href{http://dx.doi.org/10.1016/0550-3213(94)00352-1}{{\em Nucl.
  Phys.} {\bf B429} (1994)  375--388},
\href{http://arxiv.org/abs/hep-th/9403141}{{\tt arXiv:hep-th/9403141}}.

\bibitem{ZZ}
A.~Zamolodchikov and Al.~Zamolodchikov, ``{Liouville Field Theory on a Pseudosphere},''
\href{http://arxiv.org/abs/hep-th/0101152}{{\tt arXiv:hep-th/0101152}}.

\bibitem{Zamolodchikov:1995aa}
  A.~B.~Zamolodchikov and A.~B.~Zamolodchikov,
  Nucl.\ Phys.\  B {\bf 477}, 577 (1996)
  [arXiv:hep-th/9506136].

\bibitem{Troost}
C.~Jego and J.~Troost,
``{Notes on the Verlinde formula in nonrational conformal field theories},''
\href{http://dx.doi.org/10.1103/PhysRevD.74.106002}{{\em Phys. Rev. {\bf{D74}}, 106002 (2006)}}

\bibitem{EST}
T.~Eguchi, Y.~Sugawara and A.~Taormina, ``{Liouville Field, Modular Forms and Elliptic Genera},'' 
\href{http://arxiv.org/abs/hep-th/0611338}{{\tt arXiv:hep-th/0611338}}.

\bibitem{Eguchi}
T.~Eguchi, Y.~Sugawara and A.~Taormina, ``{Modular Forms and Elliptic Genera for ALE Spaces},'' 
\href{http://arxiv.org/abs/0803.0377}{{\tt arXiv:0803.0377 [hep-th]}}.

\bibitem{Teschner}
B.~Ponsot, and J.~Teschner, ``{Liouville bootstrap via harmonic analysis on a noncompact quantum group},'' 
\href{http://arxiv.org/abs/hep-th/9911110}{{\tt arXiv:hep-th/9911110}}.

\bibitem{Teschner:Lecture}
J.~Teschner, ``{A lecture on the Liouville vertex operators},'' {{\em Int. J. Mod. phys.}{\bf A19S2}(2004) 436-458},
\href{http://arxiv.org/abs/hep-th/0303150}{{\tt arXiv:hep-th/0303150}}.

\bibitem{Nekrasov1}
N.~A. Nekrasov, ``{Seiberg-Witten Prepotential From Instanton Counting},'' {\em
  Adv. Theor. Math. Phys.} {\bf 7} (2004)  831--864,
\href{http://arxiv.org/abs/hep-th/0206161}{{\tt arXiv:hep-th/0206161}}.

\bibitem{Nekrasov2}
N.~Nekrasov and A.~Okounkov, ``{Seiberg-Witten theory and random partitions},''
\href{http://arxiv.org/abs/hep-th/0306238}{{\tt arXiv:hep-th/0306238}}.

\bibitem{Nekrasov:2004vw}
N.~Nekrasov and S.~Shadchin, ``{ABCD of instantons},''
  \href{http://dx.doi.org/10.1007/s00220-004-1189-1}{{\em Commun. Math. Phys.}
  {\bf 252} (2004)  359--391},
\href{http://arxiv.org/abs/hep-th/0404225}{{\tt arXiv:hep-th/0404225}}.

\bibitem{Pestun}
V.~Pestun, ``{Localization of gauge theory on a four-sphere and supersymmetric
  Wilson loops},''
\href{http://arxiv.org/abs/0712.2824}{{\tt arXiv:0712.2824 [hep-th]}}.

\bibitem{Seiberg:1994aj}
N.~Seiberg and E.~Witten, ``{Monopoles, duality and chiral symmetry breaking in
  $\mathcal{N}=2$ supersymmetric QCD},''
  \href{http://dx.doi.org/10.1016/0550-3213(94)90214-3}{{\em Nucl. Phys.} {\bf
  B431} (1994)  484--550},
\href{http://arxiv.org/abs/hep-th/9408099}{{\tt arXiv:hep-th/9408099}}.

\bibitem{Verlinde}
E.~Verlinde,``{Fusion rules and modular transformations in 2D conformal field theory}''
\href{http://dx.doi.org/10.1016/0550-3213(88)90603-7}{{\em Nucl. Phys.} {\bf
  B300} (1988)  360--376}
  
\bibitem{Moore Seiberg}
G.~Moore and N.~Seiberg, ``{Classical and Quantum Conformal Field Theory},''
  \href{http://dx.doi.org/10.1007/BF01238857}{{\em Commun. Math. Phys.} {\bf
  123(2)} (1989)  177--254}

\bibitem{Hamiltonian Reduction}
M.~Bershadsky , ``{Conformal Field Theories via Hamiltonian Reduction},''
  \href{http://dx.doi.org/10.1007/BF02102729}{{\em Commun. Math. Phys.} {\bf
  139}(1) (1991)  71--82}
  
\bibitem{Ishibashi}
N.~Ishibashi , ``{Extra Observables in Gauged WZW Models},''
  \href{http://dx.doi.org/10.1016/0550-3213(92)90595-3}{{\em Nucl.Phys.} {\bf
  B379} (1992) 199--219}
  \href{http://arxiv.org/abs/hep-th/9110071}{{\tt arXiv:hep-th/9110071}}.


\bibitem{Izawa}
K.~Izawa , ``{Gauged WZNW formulation of Toda theories},''
{\em Tsukuba Superstrings} {\bf
  1990:43-52}

\bibitem{Holomorphic Factorization}
E.~Witten , ``{On Holomorphic Factorization of WZW and Coset Models},''
  \href{http://dx.doi.org/10.1007/BF02099196}{{\em Commun. Math. Phys.} {\bf
  144} (1992)  189--212}
  
\bibitem{CSW;nocompact}
E.~Witten,  ``{Quantization of Chern-Simons Gauge Theory with Complex Gauge Group},''
\href{http://dx.doi.org/10.1007/BF02099116}{{\em Commun. Math. Phys.} {\bf 137}
  (1991) 29--66}
  
\bibitem{Carlip 91}
S.~Carlip, ``{Inducing Liouville theory from topologically massive gravity},''
\href{heep://dx.doi.org/10.1016/0550-3213(91)90558-F}{{\em Nucl. Phys.} {\bf  B 362} 1991}.

\bibitem{Banados 94}
M.~Banados,  ``{Global Charges in Chern-Simons theory and the 2+1 black hole},''
\href{http://dx.doi.org/10.1016/0550-3213(91)90558-F}{{\em Nucl. Phys.} {\bf B362}
  (1991)  111--124}

\bibitem{Carlip 05}
S.~Carlip,  ``{Conformal Field Theory, (2+1)-Dimensional Gravity, and the BTZ Black Hole},''
\href{http://arxiv.org/abs/hep-th/0503022}{{\tt arXiv:hep-th/0503022}}.

\bibitem{Fjelstad Hwang}
J.~Fjelstad, S.~Hwang, ``{Equivalence of Chern-Simons gauge theory and WZNW model using a BRST symmetry},''
\href{http://arxiv.org/abs/hep-th/9906123}{{\tt arXiv:hep-th/9906123}}.

\bibitem{Sonoda:1988mf}
H.~Sonoda, ``{Sewing Conformal Field Theories},''
\href{http://dx.doi.org/10.1016/0550-3213(88)90066-1}{{\em Nucl. Phys.} {\bf
  B311} (1988)  401}.

\bibitem{PI QCS}
S.~Elitzur, G.~Moore, A.~Schwimmer and N.~Seiberg, ``{Remarks on the canonical quantization of the Chern-Simons-Witten theory},''
\href{http://dx.doi.org/10.1016/0550-3213(89)90436-7}{{\em Nucl. Phys.} {\bf B326 1}
  (1989)  108--134}.

\bibitem{Labastida 89a}
J.~Labastida and A.~Ramallo, ``{Operator Formalism for Chern-Simons Theories},''
\href{http://dx.doi.org/10.1016/0370-2693(89)91289-6}{{\em Phys. Lett.} {\bf B227 1}
  (1989)  92--102}.
  
\bibitem{Labastida 89b}
J.~Labastida and A.~Ramallo, ``{Chern-Simons Theory and Conformal Blocks},''
\href{http://dx.doi.org/10.1016/0370-2693(89)90661-8}{{\em Phys. Lett.} {\bf B228 2}
  (1989)  214--222}.
  
\bibitem{Ooguri Vafa}
H.~Ooguri and C.~Vafa, ``{Knot Invariants and Topological Strings}''
\href{http://arxiv.org/abs/hep-th/9912123}{{\tt arXiv:hep-th/9912123}}.
  
\bibitem{BLG}
  J.~Bagger and N.~Lambert,
  ``Gauge Symmetry and Supersymmetry of Multiple M2-Branes,''
  Phys.\ Rev.\  D {\bf 77}, 065008 (2008)
  [arXiv:0711.0955 [hep-th]],
  A.~Gustavsson,
  ``Algebraic structures on parallel M2-branes,''
  arXiv:0709.1260 [hep-th].

\bibitem{ABJM}
O.~Aharony, O.~Bergman, D.~Jafferis and J.~Maldacena ``{$\mathcal{N}=6$ superconformal Chern-Simons-Matter Theories,
 M2-branes and their gravity duals}, ''
 \href{http://dx.doi.org/10.1088/1126-6708/2008/10/091}{{\em JHEP} {10(2008)}{ 091}}.
\href{http://arxiv.org/abs/0806.1218}{{\tt arXiv:0806.1218 [hep-th]}}.
    
\end{thebibliography}\endgroup
\end{document}